\documentclass[twocolumn,reprint,aps,prl,superscriptaddress,floatfix]{revtex4-2}

\usepackage{graphicx}
\usepackage{amsmath,amssymb}
\usepackage{graphicx}
\usepackage{hyperref}
\usepackage{xcolor}
\usepackage{dsfont}
\usepackage{amssymb}
 \usepackage{mathrsfs}

\begin{document}

\title{Characterizing quantum precision enhancement for multiple currents in open quantum systems}

\author{Khalak Mahadeviya}
\affiliation{School of Physics, Trinity College Dublin, College Green, Dublin 2, D02 K8N4, Ireland}

\author{Sheikh Parvez Mandal}
\affiliation{Departamento de Física—CIOyN, Universidad de Murcia, E-30071 Murcia, Spain}

\author{Mahasweta Pandit}
\affiliation{Departamento de Física—CIOyN, Universidad de Murcia, E-30071 Murcia, Spain}

\author{Javier Prior}
\affiliation{Departamento de Física—CIOyN, Universidad de Murcia, E-30071 Murcia, Spain}

\author{Mark T. Mitchison}
\affiliation{Department of Physics, King’s College London, Strand, London WC2R 2LS, United Kingdom}

\affiliation{School of Physics, Trinity College Dublin, College Green, Dublin 2, D02 K8N4, Ireland}

\author{Saulo V. Moreira}
\affiliation{School of Physics, Trinity College Dublin, College Green, Dublin 2, D02 K8N4, Ireland}

\begin{abstract}

The kinetic uncertainty relation (KUR) constrains fluctuations of individual currents in classical nonequilibrium systems in terms of the dynamical activity, which quantifies the average number of transitions per unit of time.
Here, we derive a multi-current KUR (MKUR) establishing the precision limit for multiple currents generated in classical stochastic processes in the nonequilibrium steady state. We then apply our theory to identify quantum devices whose correlated current fluctuations surpass this classical limit by comparing the quantum system to a classical emulator with the same average currents in the steady state. In this way, violations of the emulator's MKUR have a clear physical meaning, i.e., that the quantum-coherent evolution reduces joint current fluctuations below what is possible for a classical Markovian system with the same energy-level structure and incoherent resources.
Since the MKUR incorporates correlations between currents, its violation pinpoints nonclassical fluctuations in parameter regimes where the single-current KUR is satisfied. We illustrate the essential and complementary role of these bounds for characterizing joint current precision with two examples: a coherently driven two-level system and an experimentally motivated model of a three-level heat engine subject to parasitic environmental couplings.

\end{abstract}

\maketitle

\textit{Introduction}.--- The discovery of precision limits for transport observables has reshaped the understanding of nonequilibrium thermodynamics at the nanoscale.  
Building upon previous developments in the field of stochastic thermodynamics, such as detailed and integral fluctuation theorems, the thermodynamic uncertainty relation (TUR) expresses a trade-off between fluctuations of any single-current observable and entropy production~\cite{Barato2015}. The kinetic uncertainty relation (KUR), in turn, limits the precision of current observables in terms of the dynamical activity, which quantifies the average number of stochastic transitions per unit time in the system~\cite{Garrahan2017, DiTerlizzi2018}. 
The TUR therefore implies that fluctuations can be decreased by increasing dissipation, while the KUR expresses the cost of reducing fluctuations in terms of increased dynamical activity.

It is well-established that stochastic currents generated in open quantum systems undergoing Markovian evolution can break these classical precision bounds. Such violations have been theoretically demonstrated in quantum transport setups~\cite{Ptaszynski2018, Agarwalla2018, Brandner2018, Liu2019, Guarnieri2019, Manzano2023, Kacper2023, Brandner2025} and nanoscale quantum machines~\cite{Kalaee2021, RignonBret2021, Singh2023, Meier2025}. This implies that current fluctuations can be suppressed beyond what is possible classically, thereby allowing for an increase in the signal-to-noise ratio (SNR) of current observables.
The TUR and KUR can therefore be utilized to determine quantum precision enhancement for \emph{individual} currents.
In this regard, in a setting with multiple currents, the violation of a given classical precision bound
for one particular current does not necessarily imply that quantum precision advantage can also be achieved for the other currents, unless they are perfectly correlated. Perfect correlations between current observables mean that they are linearly related, with the same average and fluctuations.
Thus, quantum precision enhancement is achieved for all currents once violation is verified for one of them. Indeed, strong correlations developed between two thermodynamic time-integrated currents in the three-level maser were shown to lead to the saturation of a multi-current \emph{quantum} KUR derived in Ref.~\cite{Moreira_2025}.

Typically, theoretical predictions of perfect correlations assume ideal conditions, e.g., the coupling of a pair of transitions defining a given current to a unique bath.
However, experimental imperfections often lead to cross-couplings of the transitions to different baths, which are usually referred to as \emph{parasitic couplings}. Such parasitic couplings are present, for example, in state-of-the-art implementations of the three-level maser in superconducting platforms~\cite{Aamir_2022, Sundelin_2026}.
In this way, even when theoretically predicted, perfect correlations can be easily hindered in realistic situations. 
Generally, if the currents are completely uncorrelated, their fluctuations become completely independent from each other. In turn, imperfect correlations mean that they are partially independent, thus displaying some degree of correlation. In this context, in order to provide a complete characterization of quantum precision advantage, it becomes necessary not only to investigate violations of precision bounds for each individual current, but also to consider the (dis)joint fluctuations between them, captured by their correlations.

In this Letter, we propose an approach enabling the complete characterization of quantum precision advantage for multiple currents in open quantum systems undergoing Markovian evolution. This involves the derivation of a multi-current classical KUR (MKUR) from multi-parameter metrology techniques, combined with the concept of classical equivalence of thermal machines~\cite{Gonzalez_2019}.  
Our MKUR consists of a bound on the \emph{generalized variance} of the current's covariance matrix, thereby incorporating correlations between currents. Our framework sheds light on the complementarity between 
the KUR and the MKUR to fully depict the system's quantum precision advantage landscape. 
This is illustrated through two examples, namely a coherently-driven qubit system and the three-level maser. To our knowledge, our approach provides the first multi-current classical KUR in the literature. Previous classical multidimensional TUR bounds have been derived in the context of Langevin dynamics~\cite{Dechant2019, Dechant2021} and Markov networks~\cite{Buschmann2026}.

\begin{figure}
    \centering
    \includegraphics[width=\linewidth]{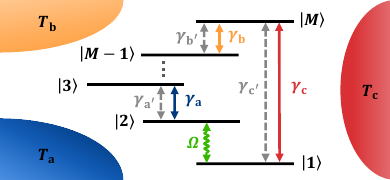}
    \caption{Sketch of a generic thermal machine with $M$ non-degenerate energy levels. The green curly arrow shows the transition coupled to a coherent drive with strength $\Omega$. The solid arrows indicate the  transitions dissipatively coupled to thermal baths at temperatures $T_\mathrm{a}$, $T_\mathrm{b}$, and $T_\mathrm{c}$, with corresponding rates $\gamma_\mathrm{a}$ (blue), $\gamma_\mathrm{b}$ (orange), and $\gamma_\mathrm{c}$ (red).  The dashed gray arrows represent parasitic cross-couplings of the transitions to other thermal baths.}
    \label{fig:M_level_schematic}
\end{figure}

\textit{Setup.}--- We consider an open quantum system with $M$ non-degenerate energy eigenstates coupled to one or more thermal baths, as illustrated in Fig.~\eqref{fig:M_level_schematic}. 
The system evolution is described by the following Markovian quantum master equation (we set $\hbar =1$),
\begin{equation}
    \label{eq:Markovian_ME}
    \dot{\rho}(t) = \mathcal{L} \rho(t) = -i \left[ H(t),\rho \right] + \sum_{k=1}^K \mathcal{D}[L_k]\rho(t),
\end{equation}
where the system Hamiltonian is $H = H_0 + V(t)$, with $H_0$ being the system's bare Hamiltonian and $V(t)$ describing the coherent coupling of the transitions between eigenstates of $H_0$.
The dissipators $\mathcal{D}[L_k]\rho = L_k\rho L_k^\dagger - \frac{1}{2}\{L_k^\dagger L_k, \rho\}$ describe the incoherent dynamics induced by the environment through the jump operators $L_k$. We assume that thermal dissipative processes described by Eq.~\eqref{eq:Markovian_ME} satisfy \emph{local detailed balance}, i.e., $L_k = e^{\Delta s_k}L_{k^\prime}^\dagger$, where $L_{k^\prime}$ represents the reversed jump transition of $L_k$, and $\Delta s_k$ is the entropy change of the environment due to a jump in channel $k$~\cite{Manzano2022}.

In the following, we consider the notion of a thermodynamic classical \emph{emulator} introduced in Ref.~\cite{Gonzalez_2019}. 
 Specifically, given a quantum system evolving according to Eq.~\eqref{eq:Markovian_ME}, its classical emulator
consists of a dissipative system that utilizes the same \emph{incoherent} resources (i.e., it couples to the same thermal baths) and has the same bare Hamiltonian $H_0$~\cite{Almanza_Marrero_2025}. In effect, the coherent part of the dynamics in Eq.~\eqref{eq:Markovian_ME} is replaced by an infinite-temperature bath, such that the same average currents as Eq.~\eqref{eq:Markovian_ME} are reproduced in the steady state without introducing additional entropy production. By denoting the transitions with the infinite-temperature bath by the jump operators $L_{k^*}$, we can write the classical master equation generating the classical emulator in the long-time limit as
\begin{equation}
    \label{eq:classical_emulator_ME}
    \dot{\rho}^\text{cl} = \mathcal{L}^\mathrm{cl} \rho^{\text{cl}}= \sum_{k^*}\mathcal{D}[L_{k^*}]\rho^\text{cl} + \sum_{k} \mathcal{D}[L_k]\rho^\text{cl}.
\end{equation}

\textit{Single-current classical KUR}.--- 
In the long-time limit, the KUR bounds the signal-to-noise ratio $D/J^2$ as~\cite{Garrahan2017, DiTerlizzi2018}
\begin{equation}
    \label{eq:classical_SKUR}
    \frac{D}{J^2} \geq \frac{1}{\dot{\mathcal{A}}},
\end{equation}
where $J$ is the average current, $D$ is the diffusion coefficient, and $\dot{\mathcal{A}}$ denotes the dynamical activity rate, which quantifies the average number of total transitions in the system per unit of time. In the context of quantum systems, there are in principle
two possible definitions for the dynamical activity rate. 
The \emph{quantum dynamical activity rate} quantifies the average rate of jumps induced by the jump operators $L_k$
in Eq.~\eqref{eq:Markovian_ME}, $\dot{\mathcal{A}}^{\mathrm{q}} = {\mathrm {Tr}}\{\sum_k L_k \rho(t)L_k^\dagger\}$~\cite{Hasegawa2020, VanVu2022}. By contrast, the \emph{classical dynamical activity rate} also takes into account the jumps $L_{k^*}$ replacing the quantum-coherent dynamics of Eq.~\eqref{eq:Markovian_ME}. It can therefore be defined as the dynamical activity of the classical emulator in Eq.~\eqref{eq:classical_emulator_ME}, i.e., $\dot{\mathcal{A}}^{\mathrm{cl}} = {\mathrm {Tr}}\{\sum_{k} L_k \rho(t)L_k^\dagger\} + {\mathrm {Tr}}\{\sum_{k^*} L_{k^*} \rho(t)L_{k^*}^\dagger\}$.
The question of whether violations of the bound with $\dot{\mathcal{A}}^{\mathrm{q}}$ represent genuine quantum enhancement has been mentioned in Ref.~\cite{Prech2025}. 
In the following, we show that the quantum dynamical activity does not always constrain the fluctuations of classical stochastic currents. Therefore, the dynamical activity entering Eq.~\eqref{eq:classical_SKUR} should always be interpreted as $\mathcal{A}^{\rm cl}$, consistent with its derivation which assumes classical stochastic dynamics~\cite{Garrahan2017, DiTerlizzi2018}.

\textit{Multi-current classical KUR}.---
The bound defined in Eq.~\eqref{eq:classical_SKUR} suffices for the characterization of quantum precision enhancement when different currents in the stationary state, e.g., exchanged with different heat baths, are perfectly correlated~\cite{Kalaee2021}.
However, for imperfectly correlated currents, their fluctuations are, to some degree, independent (or completely independent if the currents are uncorrelated).

To provide a characterization of \emph{joint} quantum precision advantage for multiple currents in an open quantum system described by Eq.~\eqref{eq:Markovian_ME}, we derive a classical MKUR on the generalized variance of \emph{two} currents generated in the corresponding classical emulation for a system with the same bare Hamiltonian and access to the same incoherent resources (heat baths).
The derivation utilizes a multi-parameter metrology approach and is detailed in the End Matter.
In the following, we briefly discuss some key definitions used in the derivation of the MKUR.
We split the set of jump channels $S = \{k \}_{k=1}^K$ into subsets of jump channels $S_\alpha \subseteq S$, with $\alpha=1,2$. This allows us to define generic counting observables $N_\alpha = \sum_{k\in S_\alpha} \nu_k N_{k}$ for some arbitrary set of scalar weights $\nu_k$, and $N_k$ is the number of jumps in channel $k$ during time $t$. Note that the set of dissipation channels $S$ excludes the channels $k^*$ representing the transitions with the infinite-temperature bath. The average currents in the stationary state are given by $J_\alpha = \mathrm{Tr}\{ \mathcal{J}_{\alpha}\rho_{\mathrm{ss}}^{\mathrm{cl}}\} = \mathrm{Tr}\{\sum_{k\in S_\alpha} \nu_k L_k\rho_{\mathrm{ss}}^\mathrm{cl}L_k^\dagger\}$, where $\mathcal{J}_{\alpha}$ are current superoperators.
As discussed in the End Matter, the covariance matrix $\Xi(N_1, N_2)$ in the long time limit can be written as $\Xi(N_1, N_2) = \mathbb{D} t$, where $\mathbb{D}$
is the diffusion coefficient matrix~\cite{Landi2024}. We denote the
diagonal elements of $\mathbb{D}$ (associated with counting observable $N_\alpha$) by $D_\alpha$. 
Just like the covariance matrix, $\mathbb{D}$ is symmetric, i.e., its off-diagonal elements satisfy $D_{\alpha\beta} = D_{\beta\alpha}$, with $\beta=1,2$, $\beta\neq\alpha$. The element $D_{\alpha\beta}$ is related to the covariance between $N_\alpha$ and $N_\beta$ as~\cite{Landi2024}
\begin{align}
    D_{\alpha\beta} = \lim_{t\to \infty} \frac{1}{t} \mathrm{Cov}(N_\alpha,N_\beta), \label{diffusionmatrix}
\end{align}
See the SI for the diffusion coefficient matrix elements' expressions.

With these definitions, we are now in a position to present our main result, which is the derivation of the following MKUR in the long-time limit
\begin{equation}
    \label{eq:Classical_MKUR}
     \frac{\det(\mathbb{D})}{J_1^2 J_2^2} = \frac{D_1D_2- {D_{12}^2}}{J_1^2 J_2^2 }  \ge \frac{(1 + \varphi_1^\text{cl})^2 (1 + \varphi_2^\text{cl})^2(1 - \varphi^{\mathrm{cl}})^2}{\dot{\mathcal{A}}_1\dot{\mathcal{A}_2}},
\end{equation}
where $\dot{\mathcal{A}}_\alpha =  \sum_{k \in S_\alpha} {\rm Tr}\{L_k \rho_\mathrm{ss}^\mathrm{cl} L_k^\dagger\}$ are \emph{partial} dynamical activities.
The corrections $\varphi_\alpha^\text{cl}$ are given by $\varphi_\alpha^\text{cl} \approx -
    \langle \mathds{1} | \mathcal{J}_\alpha {\mathcal{L}^{\mathrm{cl}}}^+\mathcal{L}_\alpha^{\mathrm{cl}} | \rho_{\text{ss}}^{\text{cl}} \rangle/J_\alpha$,
where ${\mathcal{L}^{\mathrm{cl}}}^+$ is the Drazin inverse~\cite{Landi2024} and $\mathcal{L}^{\mathrm{cl}}_\alpha \rho =  \sum_{k \in S_\alpha} \mathcal{D}[L_k]\rho$. Additionally, 
$ \varphi^{\mathrm{cl}} \approx \langle N_1^\prime \rangle \langle N_2^\prime \rangle / (1+\varphi_1^{\mathrm{cl}})\langle N_1 \rangle(1+\varphi_2^{\mathrm{cl}})\langle N_2 \rangle $, with $\langle N_\alpha \rangle = J_\alpha t $ and $\langle N_\alpha^\prime \rangle \approx -t\langle \mathds{1} | \mathcal{J}_\alpha {\mathcal{L}^{\mathrm{cl}}}^+\mathcal{L}_\beta^{\mathrm{cl}} | \rho_{\text{ss}}^{\text{cl}} \rangle$.

Note that the MKUR captures correlations between the two fluctuating currents in the long time limit through the term $D_{12}$.
When the currents are perfectly correlated, we have $\det(\mathbb{D})=0$. In this case, no additional information can be obtained through the MKUR: as the currents fluctuate together around their mean values, the single-current KUR is sufficient for the characterization of fluctuations.
We highlight that the MKUR bounds current fluctuations for a classical system with a given bare Hamiltonian and coupled to a certain number of baths.
Hence, given access to a fixed amount of incoherent resources to a quantum and a classical system with the same energy structure, the violation of the bound by the quantum system signifies that the quantum coherent evolution enables joint quantum precision enhancement for multiple currents.
In the examples below, we illustrate how the MKUR provides the characterization of quantum precision enhancement, together with the single-current KUR.

\begin{figure*}[t!]
    \centering
    \includegraphics[width=1\linewidth]{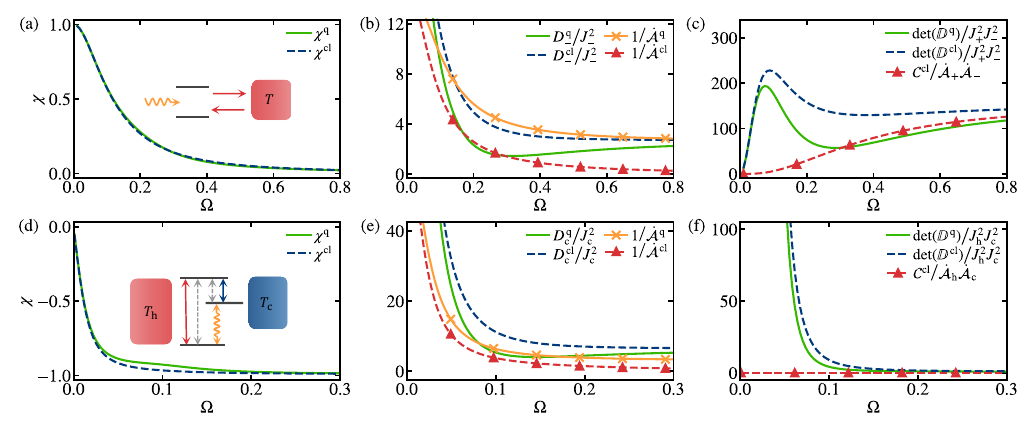}
    \caption{Characterizing quantum precision enhancement in a coherently-driven qubit (top row, illustrated in the inset of (a)) and a three-level maser (bottom row, illustrated in the inset of (d)). For the qubit, $\Delta =0$, $\gamma=0.7$ and $n = 0.05$; For the maser, $\gamma_\mathrm{h} = 0.1$, $\gamma_\mathrm{c} = 2.5$, $\gamma^\prime =\gamma^\prime_\mathrm{c} = \gamma^\prime_\mathrm{h}=0.001 $, $n_\mathrm{h} = 3.5$, $n_\mathrm{c} = 0.001$,  $n^\prime_\mathrm{h} = 4$, and $n^\prime_\mathrm{c} = 0.02$. All quantities are plotted as functions of the drive strength $\Omega$. Panels (a) and (d) show the Pearson coefficient $\chi$ for quantum (green) and classical descriptions (blue dashed). Panels (b) and (e) show relative fluctuations $D/J^2$ of counting observables $N_{-}$ (qubit) and $N_\mathrm{c}$ (maser), for quantum (green) and classical (blue dashed) descriptions, together with the single-current KUR bound $1/\dot{\mathcal A}$, evaluated using the quantum dynamical activity (orange solid, crosses) and the classical dynamical activity (red dashed, triangles). Panels (c) and (f) show scaled generalized variance $\det(\mathbb{D})/J_{1}^2 J_{2}^2$ of counting observables \{$N_+$, $N_-$\} for the qubit, and \{$N_\mathrm{h}$, $N_\mathrm{c}$\} for the maser, for quantum (green) and classical (blue dashed) descriptions, together with the classical MKUR bound $\mathcal{C}^{\mathrm{cl}}/\dot{\mathcal{A}}_{1}\dot{\mathcal{A}}_{2}$ (red dashed, triangles), where $\mathcal{C}^{\mathrm{cl}} = (1+\varphi_1^\mathrm{cl})^2(1+\varphi_2^\mathrm{cl})^2(1-\varphi^\mathrm{cl})^2$. }
    \label{fig:combined_plots}
\end{figure*}

\textit{Coherently-driven qubit}.--- We first consider a coherently driven qubit coupled to a bosonic thermal bath at temperature $T$. The system Hamiltonian is $H(t) = \omega \sigma_{+}\sigma_{-} + \Omega (\sigma_{+} e^{-i\omega_\mathrm{d} t} + \sigma_{-} e^{i\omega_\mathrm{d} t})$, where $\sigma_{x,y,z,+,-}$ are Pauli matrices, $\omega$ is the qubit transition frequency, $\omega_\mathrm{d}$ is the drive frequency, and $\Omega$ is the drive strength. The eigenstates of $\sigma_z$ are given by $\{| 0\rangle, | 1\rangle\}$. By moving to a rotating frame, we obtain a time-independent Hamiltonian, 
\begin{equation}
    \label{eq:driven_quibit_ham}
    \tilde{H} = \frac{\Delta}{2} \sigma_z + \Omega \sigma_x,
\end{equation}
with detuning $\Delta = \omega - \omega_\mathrm{d}$. 
The system dynamics is described by the following Lindblad master equation,
\begin{equation}
    \label{eq:driven_qubit_quantum_ME}
    \dot{\rho}(t) = -i[ \tilde{H},\rho(t) ] + \mathcal{D}[L_+]\rho(t) + \mathcal{D}[L_-]\rho(t).
\end{equation}
Here, $L_+ = \sqrt{\gamma n }\sigma_+$ and $L_- = \sqrt{\gamma (n+1) }\sigma_-$ are the jump operators describing thermal dissipation, with coupling strength $\gamma>0$ and bath occupation $n = \left( e^{- \omega/T }- 1\right)^{-1}$. We set $k_\mathrm{B} = 1$ throughout.
Given the set of jump channels, $S = \{+,-\}$, we define the counting observables, $N_1= N_+$ and $N_2 = N_-$, by setting the weights $\nu_+=\nu_-=1$ for $S_1 = \{+\}$ and $S_2 = \{-\}$. Therefore, we focus on the steady state unidirectional currents exchanged between the system and the bath. The average currents are given by $J_1 = J_+ =  \mathrm{Tr}\{  L_+ \rho_\mathrm{ss} L_+^\dagger \}$ and $J_{2} = J_- = \mathrm{Tr}\{  L_- \rho_\mathrm{ss} L_-^\dagger \}$.  

To fully characterize quantum precision advantage, we calculate the diffusion coefficients $D_1 = D_{-}^{\mathrm{q}}$, $D_2 = D_{+}^{\mathrm{q}}$, and $D_{12} = D_{+,-}^{\mathrm{q}}$ for the coherently-driven qubit dynamics (See the SI). We also compute $D_-^{\mathrm{cl}}$, $D_+^{\mathrm{cl}}$ and $D_{+,-}^{\mathrm{cl}}$ in the classically emulated dynamics.
In addition, we evaluate the classical MKUR bound on the right-hand side of Eq.~\eqref{eq:Classical_MKUR}, with $\dot{\mathcal{A}_1} = \dot{\mathcal{A}}_+ = \mathrm{Tr}\{  L_+ \rho_\mathrm{ss} L_+^\dagger \}  $ and $\dot{\mathcal{A}_2} = \dot{\mathcal{A}}_- = \mathrm{Tr}\{  L_- \rho_\mathrm{ss} L_-^\dagger \}  $ (See the SI for the explanation of how the correction terms $\varphi_{1}^{\mathrm{cl}}$, $\varphi_{2}^{\mathrm{cl}}$ and $\varphi^{\mathrm{cl}}$ are calculated for the classically emulated dynamics).  
In Fig.~\ref{fig:combined_plots}(a), we plot the Pearson coefficients, $-1\leq\chi^{\mathrm{q}, \mathrm{cl}}\leq 1$, $\chi^{\mathrm{q}, \mathrm{cl}} \equiv D_{+,-}^{\mathrm{q}, \mathrm{cl}}/\sqrt{D_{-}^{\mathrm{q}, \mathrm{cl}}D_{+}^{\mathrm{q}, \mathrm{cl}}}$, in the quantum and classically emulated dynamics, as a function of $\Omega$. 
We see that $\chi^{\mathrm{q}} = 1$ for $\Omega = 0$. This is because, in the absence of coherent drive, each incoherent absorption
event is followed by an incoherent emission, making the currents perfectly correlated. In turn, as $\Omega$ increases, $\chi^{\mathrm{q}}$ decreases, as incoherent emission and absorption events become increasingly uncorrelated. Note that $\chi^{\mathrm{cl}}$ closely follows the behaviour of $\chi^{\mathrm{q}}$, i.e., the infinite temperature bath hinders correlations between the currents to the same extent as the coherent drive.

In Fig.~\ref{fig:combined_plots}(b), we plot the relative fluctuations for both currents, in the quantum and classically emulated dynamics, as well as
the two versions of single-current KUR in Eq.~\eqref{eq:classical_SKUR}, i.e., for $\dot{\mathcal{A}}=\dot{\mathcal{A}}^\mathrm{q}$ and $\dot{\mathcal{A}}=\dot{\mathcal{A}^\mathrm{cl}}$. We see that the bound $1/\dot{\mathcal{A}^{\mathrm{q}}}$ is broken not only by $D_-^{\mathrm{q}}/J_-^2$, but also by $D_-^{\mathrm{cl}}/J_-^2$. The significance of this result is that the KUR using $\dot{\mathcal{A}^{\mathrm{q}}}$ does not constrain fluctuations in the classical dynamics. In contrast, as expected, $D_-^{\mathrm{cl}}/J_-^2$ satisfies the bound $1/\dot{\mathcal{A}^{\mathrm{cl}}}$. In addition, $D_-^{\mathrm{q}}/J_-^2$ violates this bound, which indicates quantum precision enhancement for this current. Conversely, $D_+^{\mathrm{q}, \mathrm{cl}}/J_+^2$ loosely satisfy both bounds, which are not shown in Fig.~\ref{fig:combined_plots}(b).
Finally, in Fig.~\ref{fig:combined_plots}(c), we plot $\det(\mathbb{D}^{\mathrm{q}})/J_+^2J_-^2$ and $\det(\mathbb{D}^{\mathrm{cl}})/J_+^2J_-^2$, as well as the classical MKUR bound in Eq.~\eqref{eq:Classical_MKUR}. While $\det(\mathbb{D}^{\mathrm{cl}})/J_+^2J_-^2$ is shown to fulfill the classical MKUR, $\det(\mathbb{D}^{\mathrm{q}})/J_+^2J_-^2$ violates it in a sustained manner for values of $\Omega$ larger than 0.35, thereby showing a robust joint quantum precision enhancement. Furthermore, we see that the violation of the MKUR occurs for small values of $\chi^{\mathrm{q}}$, i.e., when the currents are weakly correlated. This is precisely the region where the single-current KUR bound is looser in Fig.~\ref{fig:combined_plots}(b).
We can therefore see how the MKUR complements the KUR in the characterization of quantum precision advantage.

\textit{Three-level maser}.--- We consider an experimentally realizable example of a quantum thermal machine, where a three-level system is coupled to two baths and a coherent drive~\cite{Scovil1959, Geva1994}. Such a three-level engine can be implemented by symmetry-selective coupling of a superconducting artificial molecule comprising two strongly coupled transmon qubits to two microwave waveguides acting as thermal baths~\cite{Aamir_2022,Sundelin_2026}.

The time-dependent Hamiltonian of the system can be written as $H(t) = \sum_{l =1}^{3}\omega_{l}\sigma_{ll} + \Omega (\sigma_{21} e^{-i\omega_\mathrm{d} t} + \sigma_{12} e^{i\omega_\mathrm{d} t})$, with $\sigma_{jk} \equiv |j\rangle\langle k|$,  drive frequency $\omega_\mathrm{d}$, and drive strength $\Omega$. In an appropriate rotating frame, we obtain the following time-independent Hamiltonian,
\begin{equation}
    \label{eq:three_level_ham}
    \tilde{H} =- \Delta \sigma_{22} + \Omega \left(\sigma_{21}+\sigma_{12}\right),
\end{equation}
with detuning $\Delta = \omega_\mathrm{d}- \left(\omega_{2} - \omega_{1}\right)$.
Each bath primarily couples to a unique pair of transitions favored by its symmetry: the hot bath couples to $| 1\rangle \leftrightarrow |3\rangle$ and the cold bath to $| 2\rangle \leftrightarrow |3\rangle$. The corresponding jump operators are $L_{13}= \sqrt{\gamma_\mathrm{h}(n_\mathrm{h}+1)}\sigma_{13}$, $L_{31}= \sqrt{\gamma_\mathrm{h}n_\mathrm{h}}\sigma_{31}$,
$L_{23}= \sqrt{\gamma_\mathrm{c}(n_\mathrm{c}+1)}\sigma_{23}$, and $L_{32}= \sqrt{\gamma_\mathrm{c}n_\mathrm{c}}\sigma_{32}$, 
with $\gamma_\mathrm{h}$ and $\gamma_\mathrm{c}$ 
being the primary coupling rates associated with the hot and cold baths, and $n_\mathrm{h}$ and $n_\mathrm{c}$ are the corresponding thermal occupation numbers.  
Due to experimental imperfections, weak parasitic couplings of each bath to transitions with opposite symmetry are developed. In this way, the hot bath couples to $| 2\rangle \leftrightarrow |3\rangle$, with strength $\gamma^\prime_\mathrm{h}$, and the cold bath couples to $| 1\rangle \leftrightarrow |3\rangle$, with a strength $\gamma^\prime_{\mathrm c}$.
Specifically, the \emph{parasitic} jump operators are $L_{23}^\prime= \sqrt{\gamma^\prime_\mathrm{h}(n^\prime_\mathrm{h}+1)}\sigma_{23}$, $L_{32}^\prime= \sqrt{\gamma^\prime_\mathrm{h}n^\prime_\mathrm{h}}\sigma_{32}$, $L^\prime_{13}= \sqrt{\gamma^\prime_\mathrm{c}(n^\prime_\mathrm{c}+1)}\sigma_{13}$, and $L^\prime_{31}= \sqrt{\gamma^\prime_\mathrm{c}n^\prime_\mathrm{c}}\sigma_{31}$, where $n^\prime_\mathrm{h}$ and $n^\prime_\mathrm{c}$ are thermal occupations.
The Lindblad master equation describing the full system dynamics is,
\begin{align}
    \label{eq:three_level_quantum_ME}
    \dot{\rho}(t) = 
    & -i[ \tilde{H},\rho(t)] + \mathcal{D}[L_{31}]\rho(t) + \mathcal{D}[L_{13}]\rho(t) \notag\\
    &  + \mathcal{D}[L_{32}]\rho(t) + \mathcal{D}[L_{23}]\rho(t) + \mathcal{D}[L^\prime_{31}]\rho(t) \notag\\
    &+ \mathcal{D}[L^\prime_{13}]\rho(t)  + \mathcal{D}[L^\prime_{32}]\rho(t) + \mathcal{D}[L^\prime_{23}]\rho(t).
\end{align}
Note that in the absence of parasitic couplings, i.e., $\gamma^\prime_\mathrm{h} = \gamma^\prime_\mathrm{c} =0$, 
the three-level system exchanges a quantum of energy with each bath to complete an operational cycle.
This ideal case, which is widely explored in the literature, features perfectly correlated currents in the steady state~\cite{Kalaee2021, Moreira_2025}.
In contrast, the experimentally consistent model with non-zero parasitic couplings leads to multiple possible cycles of operation, which results in imperfectly correlated currents.
In the following, we consider $\gamma^\prime =\gamma^\prime_\mathrm{c} = \gamma^\prime_\mathrm{h}=0.001 $, and we denote the jump operators defining the currents of interest, or \emph{primary currents}, by $L_{\bar{h}} \equiv L_{13}$, $L_{\breve{h}} \equiv L_{31}$, $L_{\bar{c}} \equiv L_{23}$, and $L_{\breve{c}} \equiv L_{32}$. 
We define the net current observables, $N_1 = N_\mathrm{h}$ and $N_2 = N_\mathrm{c}$, by setting the weights $\nu_{\bar{h}} = \nu_{\bar{c}} = 1$ and $\nu_{\breve{h}} = \nu_{\breve{c}} = -1$, with $ S_1 = \{\bar{h},  \breve{h}\}$ and $ S_2 = \{\bar{c},  \breve{c}\} $. 
The diffusion coefficients are thus identified as $D_1 = D_{h}^{\mathrm{q}}$, $D_2 = D_{c}^{\mathrm{q}}$, and $D_{12} = D_{h,c}^{\mathrm{q}}$ (See the SI for their expressions). In turn, in the classical emulated dynamics, we have $D_1 = D_\mathrm{h}^{\mathrm{cl}}$, $D_2 = D_\mathrm{c}^{\mathrm{cl}}$, and $D_{12} = D_\mathrm{h,c}^{\mathrm{cl}}$.

In Fig.~\ref{fig:combined_plots}(d), we plot the Pearson coefficients $\chi^{\mathrm{q}, \mathrm{cl}} \equiv D_\mathrm{h,c}^{\mathrm{q}, \mathrm{cl}}/\sqrt{D_\mathrm{h}^{\mathrm{q}, \mathrm{cl}}D_\mathrm{c}^{\mathrm{q}, \mathrm{cl}}} $, which are negative because the currents are anti-correlated. We see that $\chi^{\mathrm{q}}$ and $\chi^{\mathrm{cl}}$ coincide for small $\Omega$, and are very close to each other for larger $\Omega$. They decrease monotonically with $\Omega$, eventually approaching the value of $-1$, thereby becoming almost perfectly anti-correlated. Hence, as $\Omega$ increases, both the coherent drive and infinite temperature bath eventually cancel out the decorrelating effect of the parasitic baths.
Fig.~\ref{fig:combined_plots}(e) shows that, while $D_\mathrm{c}^{\mathrm{q}}/J_\mathrm{c}^2$ violates $1/\dot{\mathcal{A}}^{\mathrm{q}}$, it does not violate $1/\dot{\mathcal{A}}^{\mathcal{\mathrm{cl}}}$, although it can closely saturate it for $\Omega \sim 0.06 $. We can also see that $D^{\mathrm{cl}}_\mathrm{c}/J_\mathrm{c}^2$ fulfills both bounds.
Lastly, Fig.~\ref{fig:combined_plots}(f) shows that $\det(\mathbb{D}^{\mathrm{q}})/J_\mathrm{h}^2J_\mathrm{c}^2$ and $\det(\mathbb{D}^{\mathrm{cl}})/J_\mathrm{h}^2J_\mathrm{c}^2$ satisfy the classical MKUR, approaching saturation as the currents become perfectly correlated. Even though we do not see violation of the MKUR, $\det(\mathbb{D}^{\mathrm{q}})/J_\mathrm{h}^2J_\mathrm{c}^2$ is always lower than $\det(\mathbb{D}^{\mathrm{cl}})/J_\mathrm{h}^2J_\mathrm{c}^2$, and sustainedly saturate the MKUR in the regime where the single-current KUR becomes increasingly loose. 

\twocolumngrid

\textit{Conclusion}.--- We derived a multi-current KUR bound on the generalized variance of multiple currents for a classical system undergoing Markovian evolution in the steady state.  
The MKUR incorporates correlations between the currents through a covariance term, thereby capturing joint current fluctuations. Our derivation used the notion of classical emulation of Markovian open quantum systems. This means that, given a quantum system with a certain bare Hamiltonian and with access to a fixed amount of incoherent resources (i.e., thermal baths), the violation of the MKUR bound, obtained for the corresponding classical emulation, indicates joint quantum precision enhancement. We illustrated the importance of the MKUR in complementing the single-current KUR for the characterization of quantum precision enhancement by considering two examples. For the coherently driven system, we found that joint quantum precision enhancement occurs precisely in the region where the KUR bound is looser, and the currents are highly independent of each other. Additionally, we showed that the version of the KUR bound with the quantum dynamical activity is violated for the classical emulated dynamics. This result supports our claim that the dynamical activity that enters the bound in Eq.~\eqref{eq:classical_SKUR} must be obtained from the classical emulated evolution, as we have done for the derivation of the MKUR. Finally, we analyzed how, in addition to the KUR, the MKUR sets the limit for quantum precision enhancement for a model of a realistic quantum heat engine, namely the three-level maser. This demonstrates how our approach can be used to identify nonclassical current fluctuations in quantum devices with correlated currents.
It would be interesting to consider the MKUR to analyze quantum precision advantage in settings involving strongly coupled systems such as coherent conductors and other quantum systems~\cite{Palmqvist2025, Maity2025, Blasi2026, Mahadeviya2026, Lucena2026, Vidal2026}.

\textit{Acknowledgments}.---We are grateful to Gonzalo Manzano for discussions that helped to inspire this project, and thank Aamir Ali, Simone Gasparinetti, and Simon Sundelin for conversations and collaboration on related topics. M.T.M. is supported by a Royal Society University
Research Fellowship. This project is co-funded by the European Union (Quantum Flagship project ASPECTS, Grant No.
101080167) and U.K. Research \& Innovation (UKRI). Views
and opinions expressed are, however, those of the authors only
and do not necessarily reflect those of the European Union,
Research Executive Agency or UKRI. Neither the European
Union nor UKRI can be held responsible for them.

\appendix

\section{End Matter: Derivation of the MKUR}

To derive the MKUR in Eq.~\eqref{eq:Classical_MKUR}, we imprint the following two parameters $\vec{\phi} = \{\phi_1,\phi_2\}$ on the monitored jump operators, as follows:
\begin{equation}
    \label{eq:parameter_imprinting}
    L_{k,\vec{\phi}} = 
    \begin{cases}        
        \sqrt{1+\phi_1} L_{k} & k\in S_1 \\
        \sqrt{1+\phi_2} L_{k} & k\in S_2
    \end{cases}. 
\end{equation}
Recall that $S_\alpha \subseteq S$ where $\alpha=1,2$ (i.e. $S_1$ and $S_2$ are disjoint subsets of the set of monitored channels $S$).
Replacing the jump operators $L_k$ in the dynamics of the classical emulator in Eq.~\eqref{eq:classical_emulator_ME} by the deformed jump operators $L_{k,\vec{\phi}}$
in Eq.~\eqref{eq:parameter_imprinting}, we can write the following master equation,

\begin{align} 
    \label{eq:classical_emulator_deformed}
    \dot\rho_{\phi_1,\phi_2}^{\mathrm{cl}}(t) &=  \sum_{k^*}\mathcal{D}[L_{k^*}]\rho^\text{cl}_{\phi_1,\phi_2} + \sum_{k \notin S_{1},S_{2}}\mathcal{D}[L_{k}]\rho^\text{cl}_{\phi_1,\phi_2}  \nonumber \\ &+(1+\phi_1) \sum_{k\in S_1} \mathcal{D}[L_k]\rho^\text{cl}_{\phi_1,\phi_2} \nonumber \\ &+  (1+\phi_2)\sum_{k\in S_2} \mathcal{D}[L_k]\rho^\text{cl}_{\phi_1,\phi_2}.
\end{align}
Note that the first two dissipators on the right-hand side are not subject to any deformation and are kept intact.
In this context, each counting observable $N_\alpha$ can be seen an estimator for each parameter $\phi_\alpha$, since $N_\alpha = N_\alpha(\gamma)$ is a function of the measurement record or trajectory $\gamma$.
In this way,  given the set of counting observables $\vec{N} = \{ N_1, N_2\}$,
the multi-parameter Cram\'er-Rao (CR) bound for the covariance matrix, $[\Xi(\vec{N})]_{ij} \equiv \langle N_i N_j\rangle - \langle N_i\rangle \langle N_j \rangle$, can be expressed as the following matrix inequality~\cite{Kay1993, Dechant2019}:
\begin{equation}\label{matrixCR}
J_{\vec{\Phi}}^T \ \Xi^{-1}(\vec{N}) \ J_{\vec{\Phi}}\leq\mathbb{F},
\end{equation}
where $ [J_{\vec{N}}]_{ij} = \partial_{\phi_j}\langle N_i \rangle $ is the Jacobian of $\langle \vec{N} \rangle$ and the Fisher information matrix $[\mathbb{F}]_{ij} \equiv F_{ij}$ is given by
\begin{align}
   F_{ij} 
  &  =  -\left\langle  \partial_{\phi_i} \partial_{\phi_j} \ln p_{\vec{\phi}}(\gamma)\right\rangle .
\end{align}
A scalar bound from Eq.~\eqref{matrixCR} can be obtained by utilizing the criteria of D-optimality~\cite{Pukelsheim2006, Suzuki2021}, i.e., by taking the determinant of both sides of the inequality. In our case, we are interested in calculating the different quantities in Eq.~\eqref{scalarCRbound} at $\phi_1 = \phi_2 = 0$, so that the derived bound provides information about the original dynamics in Eq.~\eqref{eq:classical_emulator_ME}. In this way, we have that
\begin{equation}\label{scalarCRbound}
    \frac{{\rm det(\Xi)}}{[\partial_{\phi_1}\langle N_1\rangle \partial_{\phi_2}\langle N_2\rangle -\partial_{\phi_2}\langle N_1\rangle \partial_{\phi_1}\langle N_2\rangle]^2} \Bigg|_{\vec{\phi}=0} \geq \dfrac{1}{{\rm det}(\mathbb{F})} \Bigg|_{\vec{\phi}=0} .
\end{equation}

The calculation of the elements of the Fisher information matrix at $\phi_1 = \phi_2 = 0$ yields,
\begin{align}\label{Fisherelements}
    F_{\alpha\alpha} = \mathcal{A}_\alpha, \qquad F_{\alpha\beta} =0.
\end{align}
The Fisher information is, therefore, diagonal, with its diagonal elements being equal to the partial dynamical activities $\mathcal{A}_\alpha =  \int_0^t dt^\prime\sum_{k \in S_\alpha} {\rm Tr}\{L_k \rho(t^\prime) L_k^\dagger\}$. See the SI for the derivation of Eq.~\eqref{Fisherelements}.

To calculate the terms $\partial_{\phi_\alpha} \langle N_\alpha \rangle|_{\vec{\phi}=0}$, we follow the same logic as presented in Ref.~\cite{Moreira_2025}. We start by expanding $\rho_{\phi_1,\phi_2}^{\mathrm{cl}}(t)$ for small perturbations $\phi_1, \phi_2 \ll 1$, 
\begin{equation}\label{expandedME}
\rho_{\phi_1,\phi_2}^{\mathrm{cl}}(t) \approx \rho^{\mathrm{cl}}(t) + \zeta_1\phi_1 + \zeta_2\phi_2,
\end{equation}
where $\zeta_1$ and $\zeta_2$ are both traceless.
Replacing Eq.~\eqref{expandedME} into Eq.~\eqref{eq:classical_emulator_deformed} and collecting the first order terms in $\phi_1$ and $\phi_2$ results in the following equations,
\begin{align}
    \dot{\zeta_1} = \mathcal{L}^{\mathrm{cl}}_1(\rho^{\mathrm{cl}}(t)) + \mathcal{L}^{\mathrm{cl}}(\zeta_1), \label{diffzeta1}
    \\ \dot{\zeta_2} = \mathcal{L}_2^{\mathrm{cl}}(\rho^{\mathrm{cl}}(t)) + \mathcal{L}^{\mathrm{cl}}(\zeta_2), \label{diffzeta2}
\end{align}   
where $\mathcal{L}^{\mathrm{cl}}_\alpha(\rho^{\mathrm{cl}}) =  \sum_{k \in S_\alpha} \mathcal{D}[L_{k}]\rho^{\mathrm{cl}}$.
The solution of Eqs.~\eqref{diffzeta1} and~\eqref{diffzeta2} can be written as
\begin{equation}\label{zetaalpha}
    \zeta_\alpha(t) = e^{\mathcal{L}^{\mathrm{cl}}t}[\zeta_\alpha(0)] + \int dt^\prime e^{\mathcal{L}^{\mathrm{cl}}(t-t^\prime)}\mathcal{L}^{cl}_\alpha e^{\mathcal{L}^{\mathrm{cl}}t^\prime}\rho^{\mathrm{cl}}(0).
\end{equation}
Now, we use Eq.~\eqref{expandedME} to calculate $\langle N_\alpha \rangle$ as $\langle N_\alpha \rangle \approx  \int_0^t \sum_{k \in S_\alpha} \nu_{k} (1+\phi_\alpha){\rm Tr}\{L_{k}( \rho^{\mathrm{cl}}(t) + \phi_1\zeta_1 + \phi_2\zeta_2)L_{k}^\dagger\}dt^\prime$. Taking the partial derivative, we get
\begin{align}\label{diffN_alpha}
\partial_{\phi_\alpha}\langle N_\alpha \rangle|_{\phi_\alpha = 0} &\approx \int_0^t \sum_{k 
\in S_\alpha} \nu_{k} {\rm Tr}\{L_{k} \rho^{\mathrm{cl}}(t^\prime) L_{k}^\dagger\}dt^\prime \nonumber \\ &+ \int_0^t \sum_{k \in S_\alpha} \nu_{k} {\rm Tr}\{L_{k} \zeta_\alpha(t^\prime) L_{k}^\dagger\}dt^\prime  \nonumber\\ &= \langle N_\alpha\rangle + \langle N_\alpha^\star\rangle  = (1 + \varphi_\alpha) \langle N_\alpha\rangle,
\end{align}
with  $\varphi_\alpha \equiv \langle N_\alpha^\star\rangle  / \langle N_\alpha\rangle$. Note that $\langle N_\alpha^\star\rangle$ can be expressed as 
\begin{equation}\label{Phialphastar}
    \langle N_\alpha^\star \rangle \equiv \int_0^t \sum_{k \in S_\alpha} \nu_{k} {\rm Tr}\{L_{k} \zeta_\alpha(t^\prime) L_{k}^\dagger\}dt^\prime = \int_0^t dt^\prime \ {\rm Tr}\{\mathcal{J}_\alpha\zeta_\alpha (t^\prime) \},
\end{equation}
with $\mathcal{J}_\alpha X = \sum_{k \in S_\alpha} \nu_{k} L_{k} X L_{k}^\dagger$.

We now consider the initial condition $\zeta_\alpha(0) = 0$ and replace $\zeta_\alpha(t)$ from Eq.~\eqref{zetaalpha} in the expression for $ \langle N_\alpha^\star\rangle $,
\begin{multline}\label{phistar}
    \langle N_\alpha^\star \rangle = \int_0^t dt^\prime \int_0^{t^\prime} d\tau {\rm Tr}\{\mathcal{J}_\alpha e^{\mathcal{L}^{\mathrm{cl}}(t^\prime - \tau)}\mathcal{L}^{\mathrm{cl}}_\alpha \rho(\tau) \} \\ = \int_0^{t} dt^\prime \int_0^{t^\prime} d\tau \langle \mathds{1} | \mathcal{J}_\alpha e^{\mathcal{L}^{\mathrm{cl}}(t^\prime-\tau)} \mathcal{L}^{\mathrm{cl}}_\alpha e^{\mathcal{L}^{\mathrm{cl}}\tau}|\rho(0)   \rangle, 
\end{multline}
In the equation above, we changed to the vectorization notation, i.e., $|\rho\rangle$ is the vectorized density matrix and $|\mathds{1}\rangle$ is the identity, such that ${\rm Tr}\{ \rho \} = \langle \mathds{1} | \rho \rangle$~\cite{Landi2024}. We are then able to show that Eq.~\eqref{phistar} can be approximated as
\begin{align}\label{Nstar_c}
    \langle N_\alpha^\star \rangle \approx -t\langle \mathds{1} | \mathcal{J}_\alpha {\mathcal{L}^{\mathrm{cl}}}^+\mathcal{L}_\alpha^{\mathrm{cl}} | \rho_{\text{ss}}^{\text{cl}} \rangle,
\end{align}
where ${\mathcal{L}^{\mathrm{cl}}}^+$ is the Drazin inverse~\cite{Landi2024}. We refer the interested reader in the detailed steps
for the derivation of Eq.~\eqref{Nstar_c} from Eq.~\eqref{phistar} to Ref.~\cite{Moreira_2025}, as the algebraic manipulations involved in the derivation are the same. As in the steady state $\langle N_\alpha \rangle = t\sum_{k 
\in S_\alpha} \nu_{k} {\rm Tr}\{L_{k} \rho^{\mathrm{cl}}_{\mathrm{ss}} L_{k}^\dagger\} = t J_\alpha$, 
we have that
\begin{align}\label{result_varphi_alpha}
    \varphi_\alpha^\mathrm{cl} = -\frac{\langle \mathds{1} | \mathcal{J}_\alpha {\mathcal{L}^{\mathrm{cl}}}^+\mathcal{L}_\alpha^{\mathrm{cl}} | \rho_{\text{ss}}^{\text{cl}} \rangle}{J_\alpha}.
\end{align}

Additionally, we can calculate the terms $\partial_\beta \langle N_\alpha \rangle$ in Eq.~\eqref{scalarCRbound} as
\begin{align}\label{diffN_alphabeta}
\partial_{\phi_\beta}\langle N_\alpha \rangle|_{\vec{\phi} = 0} &\approx \int_0^t \sum_{k \in S_\alpha} \nu_{k} {\rm Tr}\{L_{k} \zeta_\beta(t^\prime) L_{k}^\dagger\}dt^\prime    \equiv \langle N_\alpha^\prime \rangle.
\end{align}
Following similar steps as above, we can show that
\begin{align}\label{Nprime_c}
    \langle N_\alpha^\prime \rangle \approx -t\langle \mathds{1} | \mathcal{J}_\alpha {\mathcal{L}^{\mathrm{cl}}}^+\mathcal{L}_\beta^{\mathrm{cl}} | \rho_{\text{ss}}^{\text{cl}} \rangle.
\end{align}

We can now use the results in Eqs.~\eqref{Fisherelements},~\eqref{diffN_alpha},~\eqref{result_varphi_alpha} and~\eqref{Nprime_c} in \eqref{scalarCRbound} to write down the resulting MKUR bound,
\begin{align}
    \frac{{\mathrm{Var}(N_1)\mathrm{Var}(N_2)} - {\mathrm{Cov}}(N_1, N_2)^2}{N_1^2 N_2^2 } \notag \\  \ge \frac{(1 + \varphi_1^\text{cl})^2 (1 + \varphi_2^\text{cl})^2 (1 - \varphi^{\mathrm{cl}})^2}{\mathcal{A}_1\mathcal{A}_2},
\end{align}
where $ \varphi^{\mathrm{cl}} \equiv \langle N_1^\prime \rangle \langle N_2^\prime \rangle / (1+\varphi_1^{\mathrm{cl}})\langle N_1 \rangle(1+\varphi_2^{\mathrm{cl}})\langle N_2 \rangle $.

Finally, we can rewrite the MKUR as follows, by noting that in the long-time limit the elements of the covariance matrix can be written in terms of the corresponding diffusion coefficients, i.e., ${\rm Var}(N_\alpha)=D_\alpha t$ and ${\rm Cov}(N_\alpha, N_\beta)=D_{\alpha\beta}t$~\cite{Landi2024}, \begin{align}
    \frac{D_1D_2- {D_{12}^2}}{J_1^2 J_2^2 }  \ge \frac{(1 + \varphi_1^\text{cl})^2 (1 + \varphi_2^\text{cl})^2(1 - \varphi^{\mathrm{cl}})^2}{\dot{\mathcal{A}}_1\dot{\mathcal{A}_2}},
\end{align}
which is precisely Eq.~\eqref{eq:Classical_MKUR} in the main text.

\bibliography{references}

\newpage
\widetext
\begin{center}
	\textbf{\large Supplemental information} 
\end{center}
\setcounter{equation}{0}
\setcounter{figure}{0}
\setcounter{table}{0}
\setcounter{page}{1}
\makeatletter
\renewcommand{\theequation}{S\arabic{equation}}
\renewcommand{\thefigure}{S\arabic{figure}}


\setcounter{equation}{0}
\setcounter{figure}{0}
\setcounter{table}{0}
\setcounter{page}{1}
\setcounter{section}{0}
\makeatletter
\renewcommand{\theequation}{S\arabic{equation}}
\renewcommand{\thefigure}{S\arabic{figure}}

In this supplemental information, we first provide the derivations of the Fisher information matrix elements. Next, we present the elements of the diffusion coefficient matrix and discuss the classical emulation for the examples considered in the main text.

\section{Fisher information matrix elements}\label{FIelements}

To derive the elements if the Fisher information matrix in Eq.~\eqref{Fisherelements}, we start by considering the simplest possible system, i.e., a two-level system coupled to a single bath. The system dynamics is given by Eq.~\eqref{eq:classical_emulator_ME}, with $K=2$ and the following jump operators: $L_1 = \sqrt{\gamma_{01}} \sigma_+$ and $L_2 = \sqrt{\gamma_{10}}\sigma_-$, where $\gamma_{01}, \gamma_{10}$ are transition rates.
Additionally, the jump operators $L_{k^*}$ are written as $L_{1^*} = \sqrt{\tilde{\gamma}}\sigma_+$ and $L_{2^*} = \sqrt{\tilde{\gamma}}\sigma_-$. From this, we can write the time evolution for the system populations $p_n$ with $n=0,1$,
\begin{align}
    \dot{p}_0 = (\gamma_{10} + \tilde{\gamma}) p_1 - \Gamma_0 p_0, \\
    \dot{p}_1 = (\gamma_{01} + \tilde{\gamma}) p_0 - \Gamma_1 p_1.
\end{align}
Here, the \emph{escape rates} $\Gamma_0, \Gamma_1$ are given  $\Gamma_0 = \gamma_{01} + \tilde{\gamma}$ and $\Gamma_1 = \gamma_{10} + \tilde{\gamma}$.

Consider now a given trajectory $\gamma$ followed by the system during time $t$, 
\begin{align}
    \gamma = \{ (t_1,k_1), (t_2,k_2), \dots, (t_N, k_N) \}
\end{align}
where $N$ denotes the total number of stochastic jumps occurring during the trajectory, and $t_i$ represents the stochastic time of when a jump happens in a given dissipation channel $k_j$. We write $N=N_1+N_2+\tilde{N}$, where $N_1$ and $N_2$ are the number of jumps in the dissipation channels $1$ and $2$, respectively, and $\tilde{N}$ are the combined number of jumps in the channels $1^*$ and $2^*$. 
The probability of observing the trajectory $\gamma$ is given by
\begin{align}\label{trajectory_prob}
    p(\gamma) = p_0 \gamma_{01}^{N_1} \gamma_{10}^{N_2}\tilde{\gamma}^{\tilde{N}} e^{-\Gamma_0\tau_0}e^{-\Gamma_1\tau_1}.
\end{align}
Here, $p_0$ is the initial probability distribution (in our case, it is the steady state), while the stochastic times $\tau_1$ and $\tau_2$ denote the time the system remains in the state  $1$ and $2$, respectively, during the trajectory of duration $t$.
We now consider the parameter imprinting in Eq.~\eqref{eq:parameter_imprinting} using $\vec{\phi}= \{\phi_1, \phi_2\}$, where $S_1$ contains the channel $1$ and $S_2$ contains $2$.
This prescription amounts to replacing $\gamma_{01}$ and $\gamma_{10}$ by $(1+\phi_1)\gamma_{01}$ and $(1+\phi_2)\gamma_{10}$ in Eq.~\eqref{trajectory_prob}, which gives
\begin{align}\label{trajectory_prob_phi}
    p_{\vec{\phi}}(\gamma) = (1+\phi_1)^{N_1} (1+\phi_2)^{N_2} \gamma_{01}^{N_1} \gamma_{10}^{N_2}\tilde{\gamma}^{\tilde{N}} e^{-\Gamma_0(\phi_1)\tau_0}e^{-\Gamma_1(\phi_2)\tau_1},
\end{align}
where $\Gamma_0(\phi_1) = (1+\phi_1)\gamma_{01} + \tilde{\gamma}$ and $\Gamma_1(\phi_2) = (1+\phi_2)\gamma_{10} + \tilde{\gamma}$. In this way,
\begin{align}\label{log_trajectory_prob_phi}
    \ln p_{\vec{\phi}}(\gamma) = {N_1} \ln(1+\phi_1) + {N_2} \ln(1+\phi_2) -\Gamma_0(\phi_1)\tau_0 -\Gamma_1(\phi_2)\tau_1 + \ln(\gamma_{01}^{N_1} \gamma_{10}^{N_2}\tilde{\gamma}^{\tilde{N}}).
\end{align}
We can now calculate the elements of the Fisher information matrix
\begin{align}
    F_{\alpha\beta}(\vec{\phi}) & = -\left \langle \frac{\partial^2}{\partial\phi_\beta\partial\phi_\alpha} \ln p_{\vec{\phi}} (\gamma)\right\rangle =  -\left \langle \frac{\partial}{\partial\phi_\beta} \left(\frac{N_\alpha}{1+\phi_\alpha} - \frac{\partial \Gamma(\phi_\alpha)}{\partial\phi_\alpha}\right)\right\rangle = -\left \langle \frac{\partial}{\partial\phi_\beta} \frac{N_\alpha}{1+\phi_\alpha}\right\rangle,
\end{align}
where the brackets account for the average over all possible trajectories.
Therefore, at $\phi_1, \phi_2 = 0$, we have that
\begin{align}\label{F_alphabeta}
    F_{\alpha\beta}(\vec{\phi} = 0) = \langle N_\alpha \rangle \delta_{\alpha\beta} = \mathcal{A}_{\alpha} \delta_{\alpha \beta}.
\end{align}

We now discuss the generalization of the result~\eqref{F_alphabeta} for a system with $M > 2$ states, which is coupled to multiple baths, such that $K \geq 4$. 
In this case, at least one of the two subsets $S_\alpha$ will contain multiple channels.
As before, $N_\alpha$ represents the stochastic number of jumps occurring in the dissipation channels within $S_\alpha$, along a given trajectory. For a system with $M$ states, we must define $M$ escape rates $\Gamma_i$ for each state  $i=1,\dots,M$. The infinite temperature bath couples two of the states with rate $\tilde{\gamma}$.
 For the sake of generality, we allow these escape rates to depend on both parameters $\phi_1$ and $\phi_2$. 
 We recall that 
 no parameter is imprinted on the rate $\tilde{\gamma}$. In this way, given the parameter imprintings in Eq.~\eqref{eq:parameter_imprinting},
 the escape rates become linearly independent on the parameters, and can be generally written as $\Gamma_i(\phi_1,\phi_2) = (1+\phi_1)A_i +(1+\phi_2)B_i + C_i$,  where $C_i$ accounts for a possible constant contribution stemming from the coupling with the infinite temperature bath, and $A_i$ and $B_i$
are constants multiplying the pre-factors $(1+\phi_1)$ and $(1+\phi_2)$, respectively. Given this, we can rewrite Eq.~\eqref{log_trajectory_prob_phi} as
\begin{align}\label{log_trajectory_prob_phi_G}
    \ln p_{\vec{\phi}}(\gamma) = {N_1} \ln(1+\phi_1) + {N_2} \ln(1+\phi_2) -\Gamma_0(\phi_1, \phi_2)\tau_0 -\Gamma_1(\phi_1, \phi_2)\tau_1 + \mathrm{Const}.
\end{align}
The calculation of the elements of the Fisher information matrix gives
\begin{align}
    F_{\alpha\beta}(\vec{\phi}) & = -\left \langle \frac{\partial^2}{\partial\phi_\beta\partial\phi_\alpha} \ln p_{\vec{\phi}} (\gamma)\right\rangle =  -\left \langle \frac{\partial}{\partial\phi_\beta} \left(\frac{N_\alpha}{1+\phi_\alpha} - \frac{\partial \Gamma(\phi_\alpha, \phi_\beta)}{\partial\phi_\alpha}\right)\right\rangle.
\end{align}
At $\phi_1, \phi_2 = 0$, we therefore get $F_{\alpha\beta}(\vec{\phi} = 0) = \langle N_\alpha \rangle \delta_{\alpha\beta} = \mathcal{A}_{\alpha} \delta_{\alpha \beta}$.

\section{Diffusion coefficient matrix elements}\label{Diffusion_coeff}

The elements of the diffusion or noise coefficient matrix $\mathbb{D}$ defined in Eq.~(\ref{diffusionmatrix}) follow directly from the generator of the dynamics by means of full counting statistics~\cite{Landi2024}. In the following, we first consider the calculation of these elements in the quantum dynamics described by Eq.~\eqref{eq:Markovian_ME}.
Imprinting a counting field $\chi_\alpha$ on each counting observable $N_\alpha$ tilts the jump terms of the monitored channels in Eq.~\eqref{eq:Markovian_ME}, we get
\begin{equation}
  \mathcal{L}_{\vec{\chi}}\,\rho = -i[H,\rho]
  + \sum_{k} \left( e^{\,i \nu_k \chi_{\alpha(k)}}\, L_k \rho L_k^{\dagger}
  - \tfrac{1}{2}\{ L_k^{\dagger}L_k , \rho \} \right).
  \label{eq:tilted}
\end{equation}
The joint statistics of $\vec{N}=\{N_1,N_2\}$ is generated by the leading eigenvalue $\lambda_0(\vec\chi)$ of $\mathcal L_{\vec\chi}$, $\mathcal{L}_{\vec{\chi}}|\rho_{\rm ss}(\vec\chi)\rangle=\lambda_0(\vec\chi)|\rho_{\rm ss}(\vec\chi)\rangle$, whose Taylor expansion collects the steady state scaled cumulants of $\vec N$, so that the second cumulants read
\begin{equation}
  D_{\alpha\beta} = - \left.
  \partial_{\chi_\alpha}\partial_{\chi_\beta}\, \lambda_0(\vec{\chi})
  \right|_{\vec{\chi}=0} .
  \label{eq:Dscgf}
\end{equation} 
By expanding, we get,
\begin{equation}
  D_{\alpha\beta} = \delta_{\alpha\beta}\,\dot{\mathcal{A}}_\alpha
  - \langle \mathds{1} | \mathcal{J}_\alpha \mathcal{L}^{+} \mathcal{J}_\beta | \rho_{\rm ss}\rangle
  - \langle \mathds{1} | \mathcal{J}_\beta \mathcal{L}^{+} \mathcal{J}_\alpha | \rho_{\rm ss}\rangle ,
  \label{eq:Dab}
\end{equation}
where $\mathcal{J}_\alpha X = \sum_{k \in S_\alpha} \nu_k L_k X L_k^{\dagger}$ is the current superoperator in Eq.~\eqref{Phialphastar} and $\mathcal{L}^{+}$ is the Drazin inverse of the Liouvillian in Eq.~\eqref{eq:Markovian_ME}. Eq.~\eqref{eq:Dab} follows by summing the diffusion coefficients between individual jump channels given by,
\[
  D_{kq} = J_k\delta_{kq} - \langle\mathds{1}|\mathcal{L}_k\mathcal{L}^{+}\mathcal{L}_q|\rho_{\rm ss}\rangle - \langle\mathds{1}|\mathcal{L}_q\mathcal{L}^{+}\mathcal{L}_k|\rho_{\rm ss}\rangle.
\]
where $J_k=\mathrm{Tr}\{L_k\rho_{\rm ss}L_k^\dagger\}$, $\mathcal{L}_i\rho\equiv L_i\rho L_i^\dagger$, over  $k\in S_\alpha$ and $q\in S_\beta$, with corresponding weights $\nu_k$ and $\nu_q$, and using linearity of the Drazin inverse~\cite{Landi2024}. Because $S_1\cap S_2=\varnothing$, the $\delta_{kq}$ term survives only for $\alpha=\beta$, which is why the white-noise contribution $\sum_{k\in S_\alpha}\nu_k^2 J_k=\dot{\mathcal A}_\alpha$ in Eq.~\eqref{eq:Dab} is diagonal.  
The remaining two terms encode the temporal correlations between jumps and may either suppress or enhance the fluctuations. For $\alpha=\beta$, Eq.~\eqref{eq:Dab} reduces to the single current noise $D_\alpha = \dot{\mathcal{A}}_\alpha - 2\langle \mathds{1}|\mathcal{J}_\alpha \mathcal{L}^{+}\mathcal{J}_\alpha|\rho_{\rm ss}\rangle$, whereas for $\alpha \neq \beta$, the off-diagonal element $D_{\alpha\beta}$ carries the joint fluctuations entering $\det(\mathbb{D})$ in Eq.~\eqref{eq:Dab}. 

\emph{Coherently-driven qubit.}--- Here, for counting observables $N_+$ and $N_-$, $S_1 = \{+\}$, $S_2 = \{-\}$, and $\nu_+ = \nu_- = 1$, so the current superoperators are
\begin{equation}
    \label{eq:ap_dqubit_current_supop}
    \mathcal{J}_1 X = L_+ X L_+^{\dagger}  \quad \text{and} \quad \mathcal{J}_2 X = L_- X L_-^{\dagger}.
\end{equation}
Consequently, the steady state average currents are $J_1 = J_+ = \mathrm{Tr}\{\mathcal{J}_1\rho_\mathrm{ss}\}$ and $J_2 = J_- = \mathrm{Tr}\{\mathcal{J}_2\rho_\mathrm{ss}\}$. We also note that the dynamical activities $\dot{\mathcal{A}}_\pm = J_\pm$.
Denoting by $\gamma_+ = \gamma n$ and $\gamma_- = \gamma (n+1)$ the absorption and emission rates and by $\Gamma = \gamma_+ + \gamma_- = \gamma(2n+1)$ the total relaxation rate, Eq.~\eqref{eq:Dab} evaluated on resonance ($\Delta = 0$) gives
\begin{align}
  J_\pm &= \frac{\gamma_\pm \left( 4\Omega^2 + \gamma_\mp \Gamma \right)}{\Gamma^2 + 8\Omega^2} ,
  \label{eq:Jqubit}\\[4pt]
  D^{\rm q}_\pm &= J_\pm\,
  \frac{ \Gamma^2 \left( \Gamma^2 - 2\gamma_+\gamma_- \right)
        + 8\Omega^2 \left( 2\gamma_\mp^2 + 5\gamma_+\gamma_- - \gamma_\pm^2 \right)
        + 64\,\Omega^4 }{\left( \Gamma^2 + 8\Omega^2 \right)^2} ,
  \label{eq:Dqubit}\\[4pt]
  D^{\rm q}_{+,-} &= \gamma_+\gamma_-\Gamma\,
  \frac{ \Gamma^2 \left( \Gamma^2 - 2\gamma_+\gamma_- \right)
        + 32\,\Omega^2 \left(  \gamma_+\gamma_-  +\Omega^2 \right) }
       {\left( \Gamma^2 + 8\Omega^2 \right)^3} .
  \label{eq:Dcrossqubit}
\end{align}
For $\Omega \to 0$ the three coefficients collapse onto the same value, $\gamma_+\gamma_-(\Gamma^2 - 2\gamma_+\gamma_-)/\Gamma^3$, so that $\chi^{\rm q} = 1$ and $\det(\mathbb{D}^{\rm q}) = 0$: in the absence of the drive the two currents are the same stochastic variable, as discussed in the main text.
 
\emph{Three level maser.}--- Here $S_1 = \{\bar{h},\breve{h}\}$ and $S_2 = \{\bar{c},\breve{c}\}$ with $\nu_{\bar{h}} = \nu_{\bar{c}} = 1$ and $\nu_{\breve{h}} = \nu_{\breve{c}} = -1$, so that the net current superoperators read
\begin{equation}
  \mathcal{J}_1 X = L_{\bar{h}} X L_{\bar{h}}^{\dagger} - L_{\breve{h}} X L_{\breve{h}}^{\dagger} ,
  \qquad
  \mathcal{J}_2 X = L_{\bar{c}} X L_{\bar{c}}^{\dagger} - L_{\breve{c}} X L_{\breve{c}}^{\dagger} ,
  \label{eq:Jmaser}
\end{equation}
with partial dynamical activities $\dot{\mathcal{A}}_\mathrm{h} = \mathrm{Tr}\{ L_{\bar{h}}\rho_{\rm ss} L_{\bar{h}}^{\dagger} + L_{\breve{h}}\rho_{\rm ss} L_{\breve{h}}^{\dagger}\}$ and $\dot{\mathcal{A}}_c = \mathrm{Tr}\{ L_{\bar{c}}\rho_{\rm ss} L_{\bar{c}}^{\dagger} + L_{\breve{c}}\rho_{\rm ss} L_{\breve{c}}^{\dagger}\}$. Inserting Eq.~\eqref{eq:Jmaser} into Eq.~\eqref{eq:Dab} yields $D_\mathrm{h} = D_{11}$, $D_\mathrm{c} = D_{22}$ and $D_{h,c} = D_{12}$. The parasitic channels $L'_{ij}$ are not part of $S$ and therefore enter only through $\mathcal{L}$ and $\rho_{\rm ss}$, and not through $\mathcal{J}_\alpha$. 

We note to calculate the diffusion coefficient in the classically emulated dynamics in Eq.~\eqref{eq:classical_emulator_ME}, the jump operators $L_{k^{*}}$ and the parasitic channels in the three-level maser are not subject to the counting-field imprinting. By following the same steps above, it turns out that the diffusion coefficients $D^{\rm cl}_{\alpha\beta}$ of the classical-thermodynamic equivalent description can be calculated upon replacing $\mathcal{L}$ with $\mathcal{L}^{\rm cl}$ in Eq.~\eqref{eq:classical_emulator_ME} and $\rho_{\rm ss}$ with $\rho^{\rm cl}_{\rm ss}$, as detailed in the following section.

\section{Classical-thermodynamic equivalence}\label{Emulation}

In this Section, we provide details of the classical-thermodynamic equivalent description for the examples of a driven qubit and a three-level maser. Particularly, starting from a Lindblad master equation describing the quantum dynamics of both cases, we obtain the classical-equivalent master equation by following  the methods discussed in Refs.~\cite{Gonzalez_2019, Almanza_Marrero_2025}.
Finally, we compute the diffusion coefficients for the classical dynamics and the correction terms in Eq.~\eqref{eq:Classical_MKUR}.

We recall the Markovian quantum master equation given in Eq.~\eqref{eq:Markovian_ME},
\begin{equation}
    \label{eq:ap_quantum_Markovian_ME}
    \dot{\rho}(t) = \mathcal{L} \rho(t) = -i \left[ H(t),\rho \right] + \sum_k^K \mathcal{D}[L_k]\rho(t),
\end{equation}
where the system Hamiltonian
\begin{equation}
    \label{eq:ap_system_ham}
    H(t) = H_0 + V(t),
\end{equation}
consists of the bare system Hamiltonian $H_0$ and $V(t)$ corresponds to coherent coupling between the eigenstates of $H_0$.
In this way, the coherence induced in the system is due to $V(t)$.

As shown in Ref.~\cite{Almanza_Marrero_2025}, for such an open quantum system undergoing Markovian dynamics, all the average input and output energy currents in long-time limit depend on the diagonal elements of the steady state density matrix, bare system energy gaps, and dissipative jump rates. This implies that a classical equivalent description with identical input and output currents can be constructed by matching the diagonal elements in the steady state density matrix to those in the quantum case. This is achieved by introducing additional incoherent channels with appropriate jump rates replacing the coherent dynamics of the quantum description, without additional entropy production. The resulting classical dynamics is governed by the master equation in Eq.~\eqref{eq:classical_emulator_ME}
\begin{equation}
    \dot{\rho}^\text{cl} = \mathcal{L}^\mathrm{cl} \rho^{\text{cl}}= \sum_{k^*}\mathcal{D}[L_{k^*}]\rho^\text{cl} + \sum_{k} \mathcal{D}[L_k]\rho^\text{cl}.
\end{equation}
where $\rho^{\text{cl}}$ is the diagonal density matrix of the classical emulation, and the jump operators $L_{k^*}$ represent the additional incoherent channels. In the following, we derive the corresponding jump rates in the classically emulated dynamics for each example considered in the main text. To do so, we proceed by solving the equations of motion for the diagonal elements and nonvanishing coherences in the quantum case, as explained in detail in Refs.~\cite{Gonzalez_2019, Almanza_Marrero_2025}.

\emph{Coherently-driven qubit.}--- In the main text, our first example is a nonequilibrium open quantum system where a qubit is coupled to a time-dependent coherent drive and a dissipative bosonic thermal bath. The system Hamiltonian in the eigenbasis $\{|0\rangle, |1\rangle\}$ of $\sigma_z$ is given by,
\begin{equation}
    \label{eq:ap_dqubit_hamiltonian}
    H(t) = \omega\sigma_+\sigma_- + \Omega (\sigma_+ e^{-i\omega_d t }+ \sigma_+ e^{i\omega_d t }),
\end{equation}
with qubit frequency $\omega$, drive frequency $\omega_d$, and drive strength $\Omega$. 
Since we restrict our analysis to the weak coupling regime, we can move to the rotating frame  defined by $U(t)=e^{iOt}$, with $O = \omega_d \sigma_+ \sigma_- $ to obtain a time-independent Hamiltonian,
\begin{equation}
    \label{eq:ap_dqubit_rot_frame_Hamiltonian}
    \tilde{H} = \frac{\Delta}{2}\sigma_z + \Omega \sigma_x,
\end{equation}
where $\Delta = \omega - \omega_d$ is the detuning. The dynamics of the system in a rotating frame is described by a quantum master equation of Lindblad form given by
\begin{equation}
    \label{eq:ap_dqubit_quantum_ME}
    \dot{\rho}(t) = -i[ \tilde{H},\rho(t) ] + \mathcal{D}[L_+]\rho(t) + \mathcal{D}[L_-]\rho(t).
\end{equation}
This is Eq.~\eqref{eq:driven_qubit_quantum_ME} in the main text, with jump operators $L_+ = \sqrt{\gamma n}\sigma_+$ and $L_- = \sqrt{\gamma (n+1)}\sigma_-$ associated with the bath at temperature $T$ and bosonic occupation $n=\left( e^{-\omega/T}-1\right)^{-1}$, and coupling rate $\gamma$. In order to derive a classical equivalent master equation producing the same average energy currents as the quantum case, we write the equations of motion of the density matrix elements, $\rho_{jl}=\langle j|\rho|l\rangle$, for $j,l \in \{0,1\}$,
\begin{equation}
\begin{split}
    \label{eq:ap_dqubit_EoMs}
    \dot{\rho}_{00} & = \gamma n  \rho_{11} -  \gamma (n+1) \rho_{00} - 2\Omega~ \mathrm{Im}\left(\rho_{01} \right),  \\
    \dot{\rho}_{11} & = \gamma (n+1) \rho_{00} -  \gamma n \rho_{11} + 2\Omega~ \mathrm{Im}\left(\rho_{01} \right), \\
    \dot{\rho}_{01} & = -\frac{1}{2}\left((2n+1)\gamma - 2i\Delta \right)\rho_{01} - i\Omega (\rho_{11}-\rho_{00}). 
\end{split}
\end{equation}
At the steady state, $\dot{\rho}_{\mathrm{ss}}=0$, so we use $[\dot{\rho}_{\mathrm{ss}}]_{01} = 0$ to obtain the following equation for the steady state coherence,
\begin{equation}
    \label{eq:ap_dqubit_ss_coherence}
    [\rho_{\mathrm{ss}}]_{01} = \left([\rho_{\mathrm{ss}}]_{00}- [\rho_{\mathrm{ss}}]_{11}\right) \left( \frac{4  \Omega \Delta-i2\Omega(2n+1)\gamma}{4\Delta^2 + (2n+1)^2\gamma^2 }\right).
\end{equation} 
Now, using Eq.~\eqref{eq:ap_dqubit_ss_coherence} together with Eq.~\eqref{eq:ap_dqubit_EoMs},we get the following equations of motion for the populations of the classical-equivalent density matrix, which therefore reproduce the the populations of the quantum model in the steady state, 
\begin{equation}
\begin{split}
    \label{eq:ap_dqubit_classical_EoMs}
    \dot{\rho}_{00}^\mathrm{cl} &= \gamma n  \rho_{11}^\mathrm{cl} -  \gamma (n+1) \rho_{00}^\mathrm{cl} + \gamma^\mathrm{cl}  \rho_{11}^\mathrm{cl} - \gamma^\mathrm{cl} \rho_{00}^\mathrm{cl}, \\ 
    \dot{\rho}_{11}^\mathrm{cl} &= \gamma (n+1)  \rho_{00}^\mathrm{cl} -  \gamma n \rho_{11}^\mathrm{cl} + \gamma^\mathrm{cl} \rho_{00}^\mathrm{cl} - \gamma^\mathrm{cl}\rho_{11}^\mathrm{cl},
\end{split} 
\end{equation}
with the \emph{classical transition rate} $\gamma^\mathrm{cl}$
\begin{equation}
    \label{eq:classical_gamma_qubit}
    \gamma^{\mathrm{cl}} = \frac{4\Omega^2 \gamma (2 n +1)}{ 4 \Delta^2 +\gamma^2(2n+1)^2  }.
\end{equation}

The equations of motion obtained in Eq.~\eqref{eq:ap_dqubit_classical_EoMs} describe the classical thermodynamic equivalent dynamics of the driven qubit model. Equivalently, we can write the following classical master equation to concisely describe the classical equivalent dynamics,
\begin{equation}
    \label{eq:ap_dqubit_classical_ME}
    \dot{\rho}^\mathrm{cl}(t) = \mathcal{L}^\mathrm{cl} = \mathcal{D}[L_+^*]\rho^\mathrm{cl}(t) + \mathcal{D}[L_-^*]\rho^\mathrm{cl}(t) + \mathcal{D}[L_+]\rho^\mathrm{cl}(t) + \mathcal{D}[L_-]\rho^\mathrm{cl}(t).
\end{equation}
Here, we replaced the Hamiltonian term corresponding to the coherent evolution in Eq.~\ref{eq:ap_dqubit_quantum_ME} with incoherent dissipators $\mathcal{D}\left[L^*_k\right]$, for $L^*_{k}= \sqrt{\gamma^\mathrm{cl}} \sigma_k$ for $k \in \{+, -\}$. 
Note that the rate $\gamma^{\mathrm{cl}}$ can be seen as describing the coupling to an infinite temperature bath, which therefore does not produce additional entropy production.

Within this classical equivalent description, for counting observables $N_{1}=N_{+}$ and $N_2 = N_-$ corresponding to $S_{1} = \{+\}$ and $S_{2} =\{-\}$ with weights $\nu_+ = \nu_- = 1$, the classical equivalent current superoperators $\mathcal{J}_\alpha$ are the same as the quantum ones in Eq.~\eqref{eq:ap_dqubit_current_supop}. With this, for $\Delta = 0$, we get the corresponding average currents by evaluating $\mathrm{Tr}\{\mathcal{J}_\alpha \rho_{\mathrm{ss}}^{\mathrm{cl}}\}$,
\begin{equation}
    \label{eq:ap_dqubit_currents_cl}
    J_{\pm}^\mathrm{cl} = \frac{\gamma_\pm \left( 4\Omega^2 + \gamma_\mp \Gamma \right)}{\Gamma^2 + 8\Omega^2}, 
\end{equation}
which is equal to $J_{\pm}$ obtained in Eq.~\eqref{eq:Jqubit} for the quantum case, i.e. $\mathrm{Tr}\{\mathcal{J}_\alpha \rho_{\mathrm{ss}}\} = \mathrm{Tr}\{\mathcal{J}_\alpha \rho_{\mathrm{ss}}^{\mathrm{cl}}\}$ for $\alpha = 1,2$, thus verifying that steady state average currents are identical in the quantum and classical description. However, this is not the case for the diffusion coefficient matrix, i.e. $\mathds{D}^\mathrm{cl} \neq \mathds{D}^\mathrm{q}$. Therefore, we obtain the classical diffusion coefficients by replacing $\rho_\mathrm{ss}$ with $\rho_\mathrm{ss}^\mathrm{cl}$ in Eq.~\eqref{eq:Dab},
\begin{align}
        \label{eq:dqubit_D_cl}
        D^\mathrm{cl}_{\pm} & = J_\pm \frac{\Gamma^2 (\Gamma^2 -2\gamma_+\gamma_-)+8\Gamma\Omega^2(\Gamma +\gamma_\mp) + 64\Omega^4}{(\Gamma^2 +8\Omega^2 )^2} \\ 
        \label{eq:dqubit_corss_D_cl}
        D^\mathrm{cl}_{+,-} & = \gamma_+\gamma_-\Gamma\frac{\Gamma^2 (\Gamma^2 - 2\gamma_+\gamma_-) +8\Omega^2( \Gamma^2 + 4\Omega^2)}{(\Gamma^2 +8\Omega^2 )^3}.
\end{align}

To apply the MKUR in Eq.~\eqref{eq:Classical_MKUR} to the classical emulator described by Eq.~\eqref{eq:ap_dqubit_classical_ME}, we now identify the relevant correction terms. The individual-current correction factors, ($\varphi_1^\mathrm{cl}$) and ($\varphi_2^\mathrm{cl}$), are obtained from Eq.~\eqref{result_varphi_alpha} in the End Matter as
\begin{equation}
    \label{eq:dqubit_varphi_alpha_cl}
    \varphi_\alpha^\mathrm{cl} = \frac{\langle \mathds{1} | \mathcal{J}_\alpha {\mathcal{L}^{\mathrm{cl}}}^+\mathcal{L}_\alpha^{\mathrm{cl}} | \rho_{\text{ss}}^{\text{cl}} \rangle}{J_\alpha}.
\end{equation}
Here, ${\mathcal{L}^\mathrm{cl}}^+$ is the Drazin inverse of the classical Liouvillian in Eq.~\eqref{eq:ap_dqubit_classical_ME}. Additionally, from $\mathcal{L}^\mathrm{cl}_\alpha \rho^\mathrm{cl} = \sum_{k\in S_\alpha} \mathcal{D} [L_k] \rho^\mathrm{cl}$, we get $\mathcal{L}_1^\mathrm{cl} \rho^\mathrm{cl} = \mathcal{D}[L_+]\rho^\mathrm{cl}$ and $\mathcal{L}_2^\mathrm{cl} \rho^\mathrm{cl} = \mathcal{D}[L_-]\rho^\mathrm{cl}$.
The correction factor $\varphi^\mathrm{cl}$ is then given by
\begin{equation}
    \label{eq:dqubit_varphi_cl}
    \varphi^\mathrm{cl} = 
    \frac{\langle \mathds{1} | \mathcal{J}_1 {\mathcal{L}^{\mathrm{cl}}}^+\mathcal{L}_2^{\mathrm{cl}} | \rho_{\text{ss}}^{\text{cl}} \rangle}{(1+\varphi^\mathrm{cl}_1) J_1}
    \frac{\langle \mathds{1} | \mathcal{J}_2 {\mathcal{L}^{\mathrm{cl}}}^+\mathcal{L}_1^{\mathrm{cl}} | \rho_{\text{ss}}^{\text{cl}} \rangle}{(1+\varphi^\mathrm{cl}_2)J_2}.
\end{equation}
Using these correction factors, we numerically evaluate the MKUR for the classical emulator, obtaining the results shown in Fig.~\ref{fig:combined_plots}(c).

\emph{Three-level maser.}--- 
Our second example is an experimentally relevant model of the three-level maser thermal machine, where a three-level system interacts with a coherent drive and two thermal baths. We recall the system Hamiltonian in the main text,
\begin{equation}
    \label{eq:ap_maser_Hamiltonian}
    H(t) = \sum_{l =1}^{3}\omega_{l}\sigma_{ll} + \Omega (\sigma_{21} e^{-i\omega_\mathrm{d} t} + \sigma_{12} e^{i\omega_\mathrm{d} t}),
\end{equation}
where $\sigma_{jk} \equiv |j\rangle\langle k|$, the drive frequency is $\omega_\mathrm{d}$, and the  drive strength is $\Omega$. Similarly to the previous example, we move to the rotating frame defined by $U(t)=e^{iOt}$, with $O = \omega_1 \sigma_{11}+ (\omega_1 +\omega_d) \sigma_{22} + \omega_3 \sigma_{33}$, to get the time independent Hamiltonian
\begin{equation}
    \tilde{H} =- \Delta \sigma_{22} + \Omega \left(\sigma_{21}+\sigma_{12}\right),
\end{equation}
with $\Delta = \omega_d - (\omega_2-\omega_1)$. 
As given in Eq.~\eqref{eq:three_level_quantum_ME} in the main text, the Lindblad master equation describing the dynamics of the three-level quantum thermal machine is,
\begin{align}
    \dot{\rho}(t) = 
     & -i[ \tilde{H},\rho(t)] + \mathcal{D}[L_{31}]\rho(t) + \mathcal{D}[L_{13}]\rho(t) + \mathcal{D}[L_{32}]\rho(t) + \mathcal{D}[L_{23}]\rho(t)   \notag\\
    & + \mathcal{D}[L^\prime_{31}]\rho(t) + \mathcal{D}[L^\prime_{13}]\rho(t)  + \mathcal{D}[L^\prime_{32}]\rho(t) + \mathcal{D}[L^\prime_{23}]\rho(t).
\end{align}
This master equation include dissipators corresponding to the coupling of transitions $|1\rangle \leftrightarrow |3\rangle$ to the hot bath at temperature $T_\mathrm{h}$, with corresponding jump operators $L_{31}=\sqrt{\gamma_\mathrm{h}n_\mathrm{h}}\sigma_{31}$ and $L_{13}=\sqrt{\gamma_\mathrm{h}(n_\mathrm{h}+1)}\sigma_{13}$, where $\gamma_\mathrm{h}$ id the coupling strength and $n_\mathrm{h} =\left( e^{-(\omega_{3}-\omega_1)/T_\mathrm{h}}-1\right)^{-1}$ is the thermal occupation.
Dissipators with jump operators $L_{32}=\sqrt{\gamma_\mathrm{c}n_\mathrm{c}}\sigma_{32}$ and $L_{23}=\sqrt{\gamma_\mathrm{c}(n_\mathrm{c}+1)}\sigma_{23}$ correspond to the coupling of transitions $|2\rangle \leftrightarrow |3\rangle$ to the cold bath at temperature $T_\mathrm{c}$ with thermal occupation $n_\mathrm{c}= ( e^{-(\omega_{3}-\omega_2)/T_\mathrm{c}}-1)^{-1}$), $\gamma_\mathrm{c}$ denoting the coupling rate.

In addition to these \emph{primary} dissipation channels resulting in a unique operating cycle of the thermal machine, the primed jump operators describe the experimental situation of undesirable cross-couplings of transitions $|1\rangle \leftrightarrow |3\rangle$ to the cold bath and $|2\rangle \leftrightarrow |3\rangle$ to the hot bath, at much weaker coupling rates $\gamma^\prime_\mathrm{c}$ and $\gamma^\prime_\mathrm{h}$ respectively.

Now, we proceed to derive the classical-thermodynamic equivalent description by following the exact same steps as in the previous example of a coherently-driven qubit. We start by writing the equations of motion for the following density matrix elements, which correspond to the energy levels which are coupled to the coherent drive,
\begin{equation}
\begin{split}
    \label{eq:ap_maser_EoMs}
    \dot{\rho}_{11} & = (\gamma_\mathrm{h} n_\mathrm{h} + \gamma^\prime_\mathrm{c} n^\prime_\mathrm{c} ) \rho_{33} -  (\gamma_\mathrm{h} (n_\mathrm{h}+1) + \gamma^\prime_\mathrm{c} (n^\prime_\mathrm{c} +1)) \rho_{11} - 2\Omega~ \mathrm{Im}\left(\rho_{12} \right),  \\
    \dot{\rho}_{22} & = (\gamma_\mathrm{c} n_\mathrm{c} + \gamma^\prime_\mathrm{h} n^\prime_\mathrm{h} ) \rho_{33} -  (\gamma_\mathrm{c} (n_\mathrm{c}+1) + \gamma^\prime_\mathrm{h} (n^\prime_\mathrm{h} +1)) \rho_{22} + 2\Omega~ \mathrm{Im}\left(\rho_{12} \right), \\
    \dot{\rho}_{12} & = -\frac{1}{2}(
    \gamma_\mathrm{h} n_\mathrm{h}  + \gamma_\mathrm{c} n_\mathrm{c} + \gamma^\prime_\mathrm{h} n^\prime_\mathrm{h} + \gamma^\prime_\mathrm{c} n^\prime_\mathrm{c}
    - 2i\Delta )\rho_{12} - i\Omega (\rho_{22}-\rho_{11}). 
\end{split}
\end{equation}
In the steady state, $[\dot{\rho}_{\mathrm{ss}}]_{12} = 0$. Using Eq.~\eqref{eq:ap_maser_EoMs}, we find that
\begin{equation}
    \label{eq:ap_maser_ss_coherence}
    [{\rho}_{\mathrm{ss}}]_{12} = \left([{\rho}_{\mathrm{ss}}]_{11}- [{\rho}_{\mathrm{ss}}]_{22}\right) \left( \frac{4  \Omega \Delta-i2\Omega\left( \sum_{j \in \{h,c\}} \gamma_j n_j + \gamma^\prime_j n^\prime_j\right)}{4\Delta^2 + \left( \sum_{j \in \{h,c\}} \gamma_j n_j + \gamma^\prime_j n^\prime_j\right)^2 }\right).
\end{equation}

Finally, we can write the classically emulated dynamics for the populations of the energy levels $|1\rangle$ and $|2\rangle$, using Eq.~\eqref{eq:ap_maser_ss_coherence} together with Eq.~\eqref{eq:ap_maser_EoMs}, as follows,
\begin{equation}
\begin{split}
    \label{eq:ap_maser_EoMs}
    \dot{\rho}^\mathrm{cl}_{11} & = (\gamma_\mathrm{h} n_\mathrm{h} + \gamma^\prime_\mathrm{c} n^\prime_\mathrm{c} ) \rho^\mathrm{cl}_{33} -  (\gamma_\mathrm{h} (n_\mathrm{h}+1) + \gamma^\prime_\mathrm{c} (n^\prime_\mathrm{c} +1)) \rho^\mathrm{cl}_{11} + \gamma^\mathrm{cl}  \rho_{22}^\mathrm{cl} - \gamma^\mathrm{cl} \rho_{11}^\mathrm{cl} ,  \\
    \dot{\rho}^\mathrm{cl}_{22} & = (\gamma_\mathrm{c} n_\mathrm{c} + \gamma^\prime_\mathrm{h} n^\prime_\mathrm{h} ) \rho^\mathrm{cl}_{33} -  (\gamma_\mathrm{c} (n_\mathrm{c}+1) + \gamma^\prime_\mathrm{h} (n^\prime_\mathrm{h} +1)) \rho^\mathrm{cl}_{22} + \gamma^\mathrm{cl}  \rho_{11}^\mathrm{cl} - \gamma^\mathrm{cl} \rho_{22}^\mathrm{cl},
\end{split}
\end{equation}
with the following classical transition rate
\begin{equation}
    \label{eq:classical_gamma_maser}
    \gamma^{\mathrm{cl}} = 
    \frac{4\Omega^2\left( \sum_{j \in \{h,c\}} \gamma_j n_j + \gamma^\prime_j n^\prime_j\right)}{4\Delta^2 + \left( \sum_{j \in \{h,c\}} \gamma_j n_j + \gamma^\prime_j n^\prime_j\right)^2}.
\end{equation}
In this way, to account for the effect of coherence in the populations, the classical equivalent equations of motion incorporate additional incoherent jumps between the energy levels $|1\rangle$ and $|2\rangle$ mediated by the classical rate $\gamma^\mathrm{cl}$. Eq.~\ref{eq:ap_maser_EoMs}, together with the equation of motion of the population unaffected by the drive, can be compactly written as the following Lindblad master equation describing the classical thermodynamic equivalent dynamics of the three-level maser,
\begin{align}
    \label{eq:three_level_classical_ME}
    \dot{\rho}^{\mathrm{cl}}(t) = 
    &~ \mathcal{D}\left[ L^*_{12}\right] \rho^{\mathrm{cl}}(t) 
    +\mathcal{D}\left[ L^*_{21}\right]\rho^{\mathrm{cl}}(t) 
    +\mathcal{D}[L_{31}]\rho^{\mathrm{cl}}(t)
    + \mathcal{D}[L_{13}]\rho^{\mathrm{cl}}(t)
    + \mathcal{D}[L_{32}]\rho^{\mathrm{cl}}(t)
    \notag\\
    &+ \mathcal{D}[L_{23}]\rho^{\mathrm{cl}}(t) + \mathcal{D}[L^\prime_{31}]\rho^{\mathrm{cl}}(t) + \mathcal{D}[L^\prime_{13}]\rho^{\mathrm{cl}}(t) + \mathcal{D}[L^\prime_{32}]\rho^{\mathrm{cl}}(t)
     + \mathcal{D}[L^\prime_{23}]\rho^{\mathrm{cl}}(t),
\end{align}
with \emph{classical equivalent} jump operators $L^{*}_{12} = \sqrt{\gamma^{\mathrm{cl}}}\sigma_{12}$ and $L^{*}_{21} = \sqrt{\gamma^{\mathrm{cl}}}\sigma_{21}$. 

For this case, we evaluate the MKUR bound for net current observables $N_1 = N_\mathrm{h}$ and $N_2 = N_\mathrm{c}$ by setting the weights $\nu_{\bar{h}} = \nu_{\bar{c}} = 1$ and $\nu_{\breve{h}} = \nu_{\breve{c}} = -1$, with $ S_1 = \{\bar{h},  \breve{h}\}$ and $ S_2 = \{\bar{c},  \breve{c}\} $, denoting by $L_{\bar{h}} \equiv L_{13}$, $L_{\breve{h}} \equiv L_{31}$, $L_{\bar{c}} \equiv L_{23}$, and $L_{\breve{c}} \equiv L_{32}$ the jump operators defining the \emph{primary currents} associated with the hot and the cold baths. We then follow the same procedure as for the coherently-driven qubit to evaluate the relevant quantities entering the MKUR, using the classical dynamics above.

\end{document}